\documentclass[12pt,fleqn]{article}
\input epsf

\def\eV{{\rm eV}}

\begin{document}
\title{GZK cutoff distortion due to the energy error distribution
shape}
\author{Ivone F.\ M.\ Albuquerque\\
Instituto F\'{i}sica, Universidade de S\~{a}o\ Paulo, SP, Brazil\\ and
Space Science Laboratory, 
        University of California, Berkeley, CA 94720\\
and \\
George F.\ Smoot\\
Lawrence Berkeley National Laboratory, Space Sciences Laboratory\\ 
and Department of Physics, University of California, Berkeley, CA 94720.\\
\\
{\small Electronic mails: IFAlbuquerque@lbl.gov;  GFSmoot@lbl.gov}}

\date{\today}
\maketitle

\baselineskip=20 pt
\begin{abstract}
The development of an ultra high energy air shower has an intrinsic energy 
fluctuation due both to the first interaction point and to the cascade
development. Here we show that for a given primary energy this air shower
energy fluctuation has a lognormal distribution and thus observations
will estimate that primary energy with a lognormal error distribution.
We analyze the UHECR energy
spectrum convolved with the lognormal energy error and demonstrate
that the shape of the error distribution will interfere significantly
with the ability to observe features in the spectrum. If the standard 
deviation of the lognormal error
distribution is equal or larger than 0.25, both the shape and the
normalization of the measured energy spectra will be modified significantly.
As a consequence the GZK cutoff might be sufficiently smeared as not to 
be seen (without very high statistics). 
This result is independent of the power law of the cosmological flux.
As a conclusion we show that in order to establish the presence or not 
of the GZK feature, not only more data are needed but also that the shape of 
the energy error distribution has to be known well. The high energy tail 
and the sigma 
of the approximate lognormal distribution of the error in estimating 
the energy must be at the minimum set by the physics of showers.\\
PACs 96.40.De,96.40.Pq
\end{abstract}

\section{Introduction} 
Detection of cosmic rays arriving at the Earth with energies above 
$10^{20} $ eV questions the presence of the GZK cutoff~\cite{gzk}.
This cutoff determines the energy where the cosmic ray spectrum
is expected to abruptly steepen. Cosmic rays with ultra high energies (above 
$\sim 5 \times 10^{19} $ eV) lose energy through photoproduction of pions
when transversing the Cosmic Microwave Background Radiation (CMB). As
the CMB attenuates ultra high energy cosmic rays (UHECR) on a 50 Mpc
scale (or characteristic distance) at $10^{20}$ eV,
one can determine it's production maximum distance. An event
of $10^{20} \eV$ has to be produced within $\sim 100$ Mpc, unless it is
a non standard particle~\cite{cfk,afk}.
The absence of any powerful source located within this 
range~\cite{sommers} --- that could accelerate a cosmic ray to such an 
energy --- turns the existence 
of these events into a mystery, the so called GZK puzzle. 

The results of two important cosmic ray experiments AGASA~\cite{agasa} and 
HiRes~\cite{hires} are not consistent. Not only is the energy 
spectrum measured by HiRes systematically below the one measured by AGASA, 
but also the 
HiRes spectrum steepens around $10^{20} $ eV while AGASA's spectrum flattens 
around this energy region. The steepening in the HiRes spectrum may be in 
agreement with a GZK cutoff, while AGASA's is thought not to be. 

There are many possible ways to understand this discrepancy
~\cite{demarco,stanev}. 
The Pierre 
Auger Observatory~\cite{auger} will soon have a statistically significant 
data sample and
will certainly shed light into understanding these events.

In this article we focus on the role
of the shape of the error distribution in the energy determination. 
We show that the intrinsic 
features of an air shower results in a lognormal error distribution on the
energy determination. 
The minimum standard deviation of this distribution ($\sigma$) is set by 
physical properties of the shower. 
If additional errors due to detection --
which increases the $\sigma$ -- are not kept to minimum, the end of the
energy spectrum will be smeared in a way that the GZK feature might not be seen.

Understanding the energy error is crucial in order to determine whether 
or not the GZK cutoff is present. 
A lognormal error distribution on the reconstructed primary cosmic ray
energy is to be expected due to fluctuations
both in the shower starting point as well as from the cascade development 
\cite{gaisser}. According to simulations by the AUGER collaboration 
\cite{desrep} the depth of first interaction affects the rate of development
of the particle cascade of the shower which results in a fluctuation
of about 15\% on the number of muons and about 5\% on the eletromagnetic
component. Auger also predicts that the number of muons in a proton induced 
shower increases with primary energy as E$^{0.85}$ \cite{desrep}. 
The 15\% fluctuation will then contribute as a fixed fractional error and
the fluctuation on the number of muons on the ground will be 
$N = (1 \pm 0.15) N_0 (E\,/\,E_0)^{0.85}$. Therefore one has to add a 15\% 
contribution to the error in estimating the primary energy in addition 
to the $\sqrt{N}$ error factor.
As this shower starting fluctuation error is a percentage of the 
energy it results in a lognormal error distribution.

There are mainly two ways of determining the energy: ground detectors reconstruct
the energy based on the particle density at a certain distance from the
shower core and fluorescence detectors which 
determine the energy through the shower longitudinal profile~\cite{hires}.
The longitudinal profile determines the number of particles in the shower
per depth and it is well known to have large fluctuations. 
As mentioned above, the fluctuations  arise both from the shower starting point 
as well as from the cascade  development. 
The same is expected for the  energy determination in ground detectors, 
since the particle density depends on the number of particles.

The inherent fluctuations and  resulting lognormal error distribution will affect
crucially the analysis of data collected in ground arrays since their data
sample is collected at one particular depth. It does also affect fluorescence
data but as the energy reconstruction uses the full longitudinal profile of
the shower, there is more potential information to estimate the original 
energy.

\begin{figure} 
\centering\leavevmode \epsfxsize=220pt \epsfbox{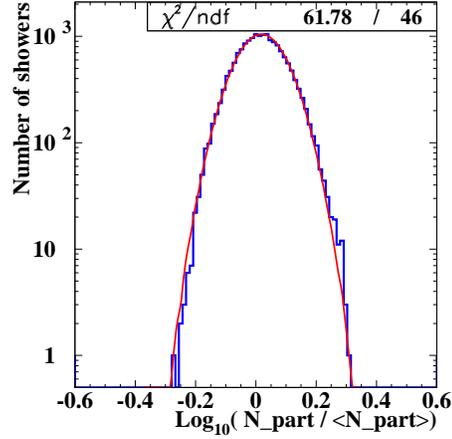}
\caption{Distribution of total number of particles at ground level.
Ratio of number of particles over average is shown.
$2\times 10^4$ showers were simulated with the Aires~\cite{aires} package.
Primary particles are $10^{20}$ eV protons and $\langle {\rm N_{part}}
\rangle = 2.7 \times 10^{10}$. Superimposed is a lognormal curve with $\sigma = 0.08$.}
\label{fig:grdp}
\end{figure}

\begin{figure} 
\centering\leavevmode \epsfxsize=220pt \epsfbox{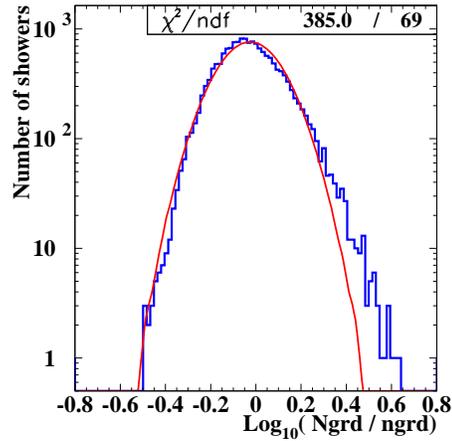}
\caption{Same as Figure~\ref{fig:grdp} but for $10^{18}$ eV protons
as the primaries. Here $\langle {\rm N_{part}} \rangle = 14.7 \times 10^7$.
Superimposed is a lognormal curve with $\sigma = 0.14$. The poor fit is due to 
an excess of simulated events relative to the lognormal at the high end.}
\label{fig:grdp18}
\end{figure}

Figure~\ref{fig:grdp} shows the distribution
of particles at ground level for $2 \times 10^4$ simulated 
showers~(using \cite{aires}) 
from $10^{20}$ eV protons. A lognormal fit with $\sigma = 0.08$ is
superimposed and it is clear that the distribution has a lognormal shape.
The same distribution for showers from $10^{18}$~eV
protons is shown in Figure~\ref{fig:grdp18}. The poor fit is due to an excess
of simulated events relative to the lognormal at the high end.
The standard deviation of the fit is 0.14. Effects due to the errors with
asymmetrical and non-gaussian tails are shown in \cite{vaz}.


\begin{figure} 
\centering\leavevmode \epsfxsize=220pt \epsfbox{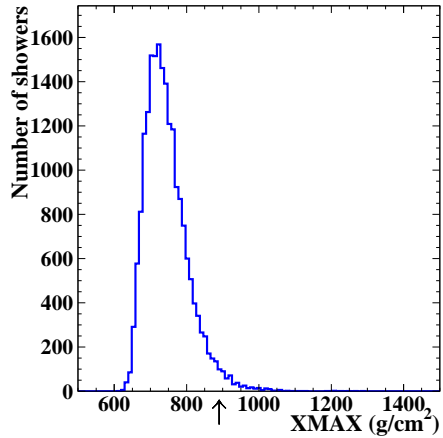}
\caption{Maximum shower depth (XMAX) distribution for $2 \times 10^{4}$ showers 
with $10^{18}$ eV protons as the primaries. The arrow in the XMAX axis indicates
where an analysis cut will be applied.}
\label{fig:xmax18}
\end{figure}

\begin{figure} 
\centering\leavevmode \epsfxsize=220pt \epsfbox{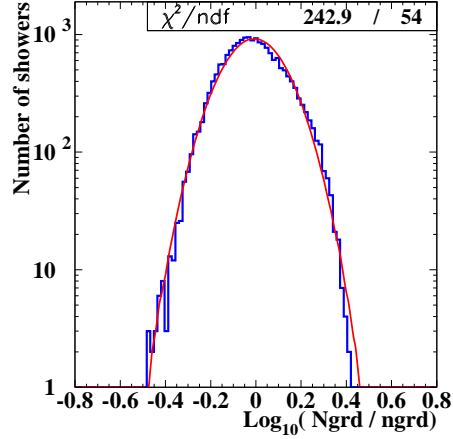}
\caption{Same as Figure~\ref{fig:grdp18} but events with XMAX $>$ 890 g/cm$^2$
were removed. Superimposed is a lognormal curve with $\sigma = 0.13$. 
The fit improves in relation to Figure~\ref{fig:grdp18}.}
\label{fig:cut18}
\end{figure}

The simulated showers used Sybill interaction model and assumed that the ground 
was at sea level (defined in Aires \cite{aires} as 0\,m or 1036\,g/cm$^2$). A more
thorough analysis is under way to understand why the error distribution for lower
energies (as in Figure~\ref{fig:grdp18}) deviates from the lognormal shape.
However it is clear that most of the events in excess come from the tail of the
maximum depth (XMAX) of the shower distribution. In Figure~\ref{fig:xmax18} we
show the XMAX distribution for the same events used in Figure~\ref{fig:grdp18}.
If we cut events with XMAX $>$ 890 g/cm$^2$ from this data set, 
the ground particles distribution will lose part of the excess events. This 
distribution is shown in Figure~\ref{fig:cut18}. 
These excess events, if included, would only make more exaggerated the effect we discuss here.


The results shown in 
Figures~\ref{fig:grdp} and~\ref{fig:grdp18} are also dependent on the location of
the ground level. We have also simulated events with ground above
sea level at 950\,g/cm$^2$. The lognormal continues to fit well the
$10^{20}$ eV distribution and its $\sigma$ improves to 0.05. The $10^{18}$ eV 
distribution still has an excess but the chisquare improves to 4 and the $\sigma$
to 0.10.

Below we will describe how we determine the UHECR spectrum
assuming a injection power spectrum from cosmologically distributed sources. 
We account for energy loss due to propagation through the CMB. 
We then describe how the energy error is
evaluated and how it affects the energy reconstruction and the determination
of the GZK cutoff.

\section{Analytical determination of UHECR propagation and energy 
spectrum}
Our analytical approach assumes a cosmological cosmic ray flux. We 
assume extragalactic sources isotropically distributed at different 
redshifts~\cite{blan}. These
sources produce a power law energy spectrum (injection spectrum) which is
assumed to be:
\begin{equation}
F(E) = k E^{-\alpha} \exp\left(-\frac{E}{E_{max}}\right)
\label{eq:flux}
\end{equation}
where $E$ is the cosmic ray energy, $k$ is a normalization factor, $\alpha$ 
is the spectral index and $E_{max}$ is the maximum energy at the source.

\begin{figure} 
\centering\leavevmode \epsfxsize=250pt \epsfbox{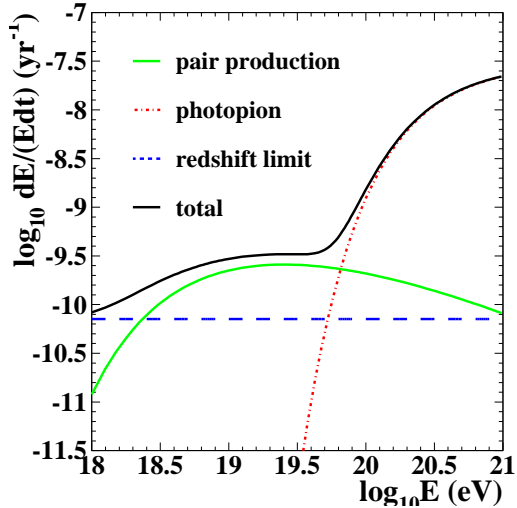}
\caption{Energy losses (as labeled) of a proton transversing the CMB at 
present epoch.}
\label{fig:enloss}
\end{figure}

The energy degradation of protons through the CMB includes losses 
due to pair production~\cite{blu,geddes}, expansion of the universe~\cite{bere} 
and photopion production~\cite{bere}. These losses at present
epoch are shown in Figure~\ref{fig:enloss}.
We include current values
for the matter and dark energy density ($\Omega_M$ and $\Omega_\Lambda$)
when considering the energy loss due to expansion of the universe 
($\beta_z$):

\begin{equation}
\beta_z(E,z) = H_0 \sqrt{\Omega_M (1 + z)^3 + \Omega_\Lambda}
\end{equation}
where $\beta$ is defined as $\beta = 1/E \times dE/dt$ and 
$\Omega_M = 0.3$, $\Omega_\Lambda = 0.7$ and $H_0 = 75 $
~km~s$^{-1}$~Mpc$^{-1}$.

Also the energy losses due to pair or to photopion production ($\beta(E,z)$)
at a certain epoch with redshift z, is corrected. Since the number density
of the cosmic background photons varies as $n = n_0 \, (1+z)^3$ the energy loss
at z differs from the energy loss today ($\beta_0(E)$) in the following way:

\begin{equation}
\beta(E,z) = (1 + z)^3 \beta_0((1+z)E)
\end{equation}

Once all energy loss mechanisms are known, the energy with which a
proton has to be generated in order to account for the energy observed
today can be determined. The generated energy depends on the distance
or epoch (redshift) from today. This can be well determined by a
modification factor $\eta(E,z)$~\cite{bere} which relates the generated
energy spectrum to the modified (and measured) one.

The cosmological flux assumes the observer in the center of a sphere
of large radius and an isotropic density of sources \cite{blan}.
The flux at the Earth is given by:
\begin{eqnarray*}
j(E) & = & \frac{c}{4 \pi H_0} \int_0^z F(E_g) 
\left(\frac{E_g}{E}\right)^{-\alpha} (1 + z)^m 
\frac{dE_g}{dE} \\
& & \times \frac{dz}
{(1 + z)\left[\Omega_M (1 + z)^3 + \Omega_\Lambda\right]^{1/2}}
\end{eqnarray*}
where $E_g$ is the generated cosmic ray energy (at a source located with
redshift $z$); $F(E)$ is given by Equation~\ref{eq:flux}; $E$ is the cosmic 
ray energy determined at the Earth; $\alpha$ is the same spectral index as
in Equation~\ref{eq:flux}; $m$ accounts for the luminosity evolution of the
sources and $c$ is the speed of light. We assume $m = 0$ and therefore do not
take luminosity evolution into account.  
The modification factor $\eta$ is given by:
\begin{equation}
\eta = \int_0^z \left(\frac{E_g}{E}\right)^{-\alpha} \frac{dE_g}{dE} \frac{dz}
{(1 + z)\left[\Omega_M (1 + z)^3 + \Omega_\Lambda\right]^{1/2}}
\end{equation}

\begin{figure} 
\centering\leavevmode \epsfxsize=250pt \epsfbox{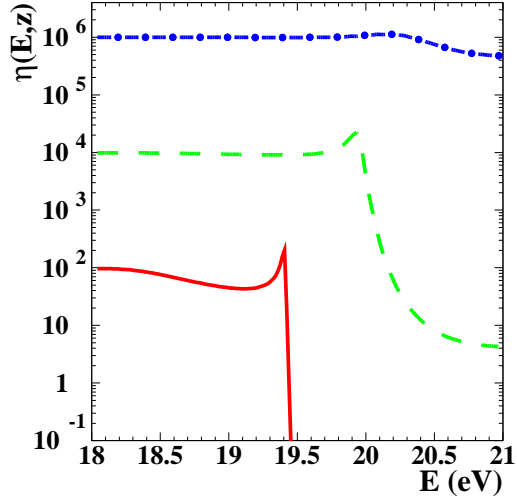}
\caption{Modification factor~\cite{bere} from our analytical calculation
versus cosmic ray energy. Curves are for different redshifts (top: 
$z = 0.002$; middle: $z = 0.02$; bottom $z = 0.2$)
and assuming no cosmological constant in order to compare results to 
\cite{bere,demarco}.}
\label{fig:mod}
\end{figure}

For comparison, we determine the modification factor for arbitrary
redshifts and assuming no cosmological constant. Our results match those
of~\cite{bere,demarco} and are shown in Figure~\ref{fig:mod}.

\begin{figure} 
\centering\leavevmode \epsfxsize=250pt \epsfbox{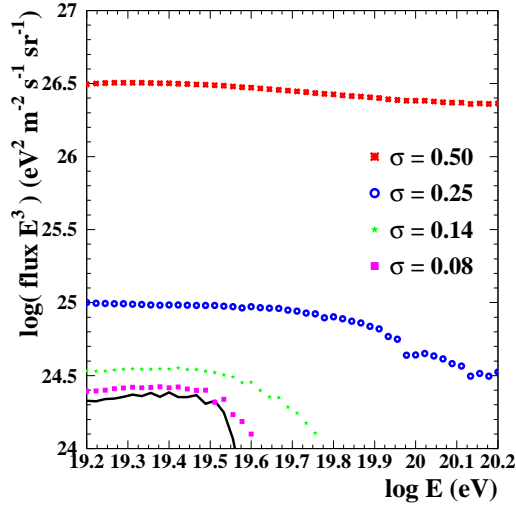}
\caption{Cosmic ray energy spectrum ($\times E^3$) from a cosmological 
flux (solid black line) with spectral index $\alpha = 2.6$. The other curves 
are the energy spectrum convoluted with a lognormal
error with standard deviations $\sigma$ as shown.}
\label{fig:specg}
\end{figure}

In Figure~\ref{fig:specg} is shown (black solid curve) the cosmic ray expected
flux versus energy ($E$) at Earth multiplied by $E^3$ 
from a cosmological injection spectrum with $\alpha = 2.6$. The expected GZK
feature is present.

\section{Errors effects on the energy reconstruction}
We will now assume that the cosmic ray energy spectrum from a cosmological
isotropic distribution of sources 
is the true spectrum. To understand how an error in the reconstructed energy
affects the spectrum, we convolute the cosmological flux assuming a lognormal
error on the energy.

The lognormal distribution is given by
\begin{equation}
\frac{dP(E',E)}{d\ln E} = k 
\exp\left[-\frac{1}{2\sigma^2} \log^2\frac{E'}{E}\right]
\end{equation}
where $k = 1\, /\, \sqrt{2 \pi}\sigma$ is a normalization to unit area and 
$\sigma$  is the standard deviation of $\log_{10}E$. 
When a lognormal error in the energy reconstruction is assumed, the flux will be 
convoluted in the following way:
\begin{equation}
dF'(E) = F(E') \frac{dP(E',E)}{dE} dE' 
\end{equation}
where F is given by Equation~\ref{eq:flux}.

The expected flux ($\times E^3$) for energies reconstructed with a lognormal
error distribution is shown in Figure~\ref{fig:specg}. The curves are for 
a spectral index
$\alpha = 2.6$ and $\sigma = 0.08$, 0.14, 0.25 and 0.5 as labeled.

It is very clear that not only the flux increases by a constant factor but
also the GZK feature is smeared. As shown in Figures~\ref{fig:grdp}
and \ref{fig:grdp18}, the standard deviation in the lognormal 
distribution will be obtained in an ideal case where thousands of events 
are detected depends on the energy of the primary particle. It is
0.08 for a $10^{20}$ eV proton and 0.14 for a $10^{18}$ eV proton.
Figure~\ref{fig:specg} shows that if the standard deviation is above
0.14 the GZK cutoff will show up at higher energies than in the true 
spectrum.

\section{Results and Conclusions}
Figure~\ref{fig:specg} shows how the energy spectrum from a cosmological
flux is smeared due to a lognormal error in the energy reconstruction.
Intrinsic shower fluctuations leads to a lognormal distribution of
observed energy deposition and number of particles in the shower.
A standard deviation of $log_{10}E$ equal to 0.25 is enough to modify not only the 
shape as
well as the normalization of the spectrum measured at the Earth.
As a consequence the GZK feature will be smeared and might not be detected 
at all. Such a $\sigma$ (standard deviation of $\log_{10}E$) can
easily result from a detector that only samples a small portion of
the total number of particles. This will be more crucial
to ground detectors since their particle sample is detected all at one 
height. The standard deviation of the intrinsic 
energy error distribution for ground detectors is expected to be larger 
than for fluorescence detectors.

The air fluorescence
detectors will have lower intrinsic lognormal standard deviation as they
observe the full development of the shower through the range of view.
They miss observing only what goes into the ground or is out of the field
of view.

As the Pierre Auger Observatory will not only increase the data sample
to a significant level, but also combine both ground and fluorescence
techniques, it will have constraints to understand and better control
the errors in the energy reconstruction. In this way it is
possible to keep the standard deviation of the intrinsic lognormal
energy error to its minimum value.

The lognormal curves shown in 
Figure~\ref{fig:grdp} and \ref{fig:grdp18} have $\sigma = 0.08$ and 0.14
respectively. However one can expect a larger value from an observed distribution
since the detectors sample only a portion of the total number of particles.
On the other hand
the standard deviation of the distribution depends on the ground level altitude
and
therefore an analysis equivalent to ours has to be done for a specific
depth.

\begin{figure} 
\centering\leavevmode \epsfxsize=300pt \epsfbox{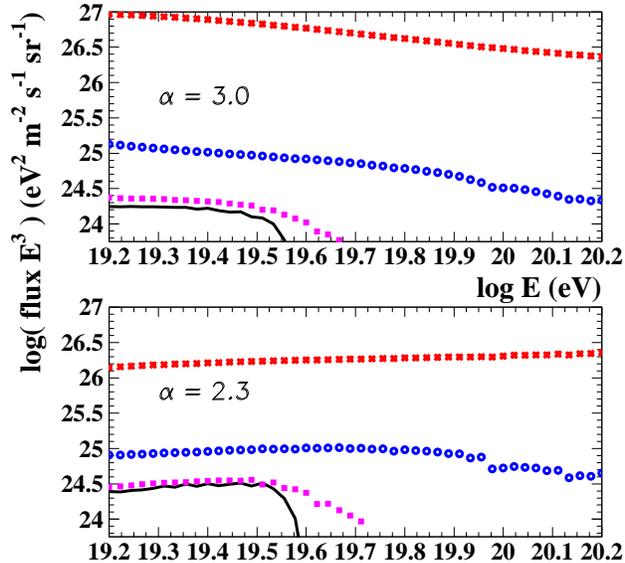}
\vspace*{-2.5cm}
\caption{Same as Figure~\ref{fig:specg} but with spectral index $\alpha = 3.0$
(top) and $\alpha = 2.3$ (bottom). Curve with crosses (x) has 
$\sigma = 0.5$; circles (o), $\sigma = 0.25$ and squares, $\sigma = 0.1$.}
\label{fig:2spec}
\end{figure}

Figure~\ref{fig:2spec} shows that the lognormal error in the energy
is also affected by the spectral index of the injection spectra. However
the error in the energy reconstruction will smear the flux in a
significant way independently of the spectral index.

We have shown that a lognormal error in the energy reconstruction of
the UHECR spectra will affect not only the shape but also the normalization
of the measured energy spectra. A standard deviation equal to or greater 
than 0.25 will smear the GZK feature. As a consequence this feature will 
not be seen. 
This result is independent of the spectral index of the injection spectra.

In conclusion, the establishment of the presence or not of the GZK cutoff in 
the 
UHECR spectrum depends not only in a larger data sample but also in the 
determination of the shape of the energy error distribution. The standard 
deviation of this distribution has to be kept to its intrinsec value. If
it is equal or greater to 0.25 the GZK feature will be smeared and not be
detected.

{\em Acknowledgements --} 
We thank Don Groom for useful comments.
This work was partially supported by NSF Grant Physics/Polar Programs
No. 0071886 and in part
by the Director, Office of Energy Research, Office of High Energy and
Nuclear Physics, Division of High Energy Physics of the U.S. Department
of Energy under Contract Num. DE-AC03-76SF00098 through the Lawrence
Berkeley National Laboratory.

\end{document}